\documentclass[%
 aip,
 amsmath,amssymb,
reprint,
]{revtex4-1}

\usepackage{graphicx}
\usepackage{dcolumn}
\usepackage{bm}
\usepackage[mathlines]{lineno}

\usepackage[utf8]{inputenc}
\usepackage[T1]{fontenc}
\usepackage{mathptmx}
\usepackage{etoolbox}
\usepackage{xcolor, soul}
\makeatletter
\def\@email#1#2{%
 \endgroup
 \patchcmd{\titleblock@produce}
  {\frontmatter@RRAPformat}
  {\frontmatter@RRAPformat{\produce@RRAP{*#1\href{mailto:#2}{#2}}}\frontmatter@RRAPformat}
  {}{}
}%
\makeatother
\begin{document}

\preprint{}

\title{Superconducting properties of lifted-off Niobium nanowires}
\author{A. Kotsovolou}%
\altaffiliation{A. Kotsovolou and F. Soofivand equally contributed to this work.}%
\affiliation{Dipartimento di Fisica ``E. Fermi", Università di Pisa, Largo Pontecorvo 3, I-56127 Pisa, Italy
}
\affiliation{INFN Sezione di Pisa, Largo Bruno Pontecorvo 3, I-56127, Pisa, Italy
}%

\author{F. Soofivand}%
\altaffiliation{A. Kotsovolou and F. Soofivand equally contributed to this work.}%
\affiliation{ 
NEST, Istituto Nanoscienze-CNR and Scuola Normale Superiore, Piazza San Silvestro 12, I-56127 Pisa, Italy
}%

\author{P. Singha}
\affiliation{NEST Laboratory, Istituto Nanoscienze-CNR, Piazza San Silvestro 12, I-56127 Pisa, Italy}

\author{D. Cecca}%
\affiliation{Dipartimento di Fisica ``E. Fermi", Università di Pisa, Largo Pontecorvo 3, I-56127 Pisa, Italy
}

\author{R. Balice}%
\affiliation{Dipartimento di Fisica ``E. Fermi", Università di Pisa, Largo Pontecorvo 3, I-56127 Pisa, Italy
}

\author{F. Carillo}%
\affiliation{INFN Sezione di Pisa, Largo Bruno Pontecorvo 3, I-56127, Pisa, Italy
}%

\author{C. Puglia}%
\affiliation{INFN Sezione di Pisa, Largo Bruno Pontecorvo 3, I-56127, Pisa, Italy
}%

\author{G. De Simoni}%
\affiliation{ 
NEST, Istituto Nanoscienze-CNR and Scuola Normale Superiore, Piazza San Silvestro 12, I-56127 Pisa, Italy
}%

\author{F. Bianco}
\email{federica.bianco@nano.cnr.it}
\affiliation{NEST Laboratory, Istituto Nanoscienze-CNR, Piazza San Silvestro 12, I-56127 Pisa, Italy}

\author{F. Paolucci}%
\email{federico.paolucci@unipi.it}
\affiliation{Dipartimento di Fisica ``E. Fermi", Università di Pisa, Largo Pontecorvo 3, I-56127 Pisa, Italy
}
\affiliation{INFN Sezione di Pisa, Largo Bruno Pontecorvo 3, I-56127, Pisa, Italy
}%


\begin{abstract}
Hybrid superconductor/semiconductor devices play a crucial role in advancing quantum science and technology by merging the properties of superconductors and semiconductors. To operate these devices at high temperature, Niobium could substitute the widespread aluminum as superconducting element. Niobium devices show the best superconducting properties when shaped by etching, but this technique is often incompatible with semiconductors and two-dimensional materials. Our work investigates the influence of oxygen diffusion on the superconducting transition of Nb nanowires fabricated by lift-off technique. To this scope, we fabricate and measure Nb devices of different width ($W$) and thickness ($t$).
By using the Berezinskii-Kosterlitz-Thouless ($BKT$) model for charge transport, we demonstrate that our nanowires behave as two-dimensional superconductors regardless of $W$ and $t$. 
While the normal-state transition temperature ($T_N$) remains constant with decreasing $W$, the temperature of the fully superconducting state ($T_S$) decreases. Thus, the superconducting transition width ($\delta T_C$) increases as $W$ shrinks, due to oxygen diffusion from the lithography resist occurring during deposition. These insights provide essential knowledge for optimizing Nb-based hybrid quantum devices, paving the way for operating temperatures above 2 K and contributing to the development of next-generation quantum technologies.
\end{abstract}

\maketitle

\section{Introduction}
Hybrid superconductor/semiconductor devices are fundamental building blocks for state-of-the-art quantum science and technology. Indeed, these systems present a double-quantum nature by combining the typical characteristics of superconductivity, such as long range coherence, dissipationless charge transport, and energy confinement, together with the field-effect tunability and the unique low-dimensional electronic properties of semiconductors.
This peculiar union gives rise to topological quantum states with exotic implications to superconductivity\cite{Sato_2017}, such as p-wave symmetry, quantized heat transport through edge states, non-Abelian statistics and Majorana bound states.
From a technological point of view, these hybrid systems have been exploited in various superconducting qubits, such as fluxoniums \cite{Pita2020}, gatemons \cite{Larsen2015}, and protected qubits \cite{Larsen2020}. For instance, gatemons employ nanowires \cite{DeLange2015,Larsen2015}, two-dimensional electron gases \cite{Casparis2018-sc,Sagi2024-ej}, carbon nanotubes \cite{Riechert2025-gr} and graphene \cite{Wang2019-gv}.
Typically, these devices exploit aluminum as superconducting material, owing to its ease of fabrication, robust superconducting properties, and the formation of high-quality oxide tunnel barriers. Currently, the operation of Al-based technology is limited well under the Al critical temperature $T_{C,Al}\sim1.2$ K, thus requiring complex and expensive cryogenic systems.
Niobium is a promising candidate to push the operation of superconducting quantum technology towards higher temperatures. Indeed, Nb thin films and etched devices show a best critical temperature $T_{C,Nb}\sim9.3$ K. Etching is frequently excluded from the fabrication of two-dimensional material devices, because this process can severely damage or even completely destroy the latter. The fabrication of Nb-based hybrid quantum technologies typically follows a bottom-up approach, relying on lift-off techniques for the superconducting structures.
Generally, these devices show operating temperatures well below 2 K, as extensively shown in several Nb-graphene hybrid systems \cite{Ben_Shalom2016-ci,Thompson2017,Li_2018}. In many cases, the decreased operation temperature is attributed to the sizable contact resistance between the superconducting electrodes and the proximitized semiconductor \cite{Mayer2019}. Actually, the critical temperature of Nb thin films depends strongly on the presence of interstitial oxygen incorporated during the growth \cite{Manzo2024,Kock1974,DeSorbo1963}. Indeed, Nb devices are affected by the fabrication process and the presence of a resist mask during the film growth \cite{Strenzke2024}. As a consequence, Niobium/semiconductor devices with vanishing contact resistance show suppressed operating temperature too, limiting the practical applications of quantum technology. 

So far, the dependence of the superconducting properties of Nb nanodevices obtained by bottom-up approach on their physical dimensions has never been addressed. In this work, we demonstrate the impact of oxygen diffusion on the superconducting transition temperature of Nb nanowires of different width and thickness fabricated by lift-off process. In particular, we attribute the widening of the superconducting transition of our devices to the diffusion of oxygen from the resist during the growth process. Our study provides fundamental insights for the development of nanostructured superconductor/semiconductor quantum devices where etching of Nb films is not possible, with the potential of pushing the operation of hybrid quantum technologies to temperatures well above 2 K.

\begin{figure}
\centering
\includegraphics[width=0.95\columnwidth]{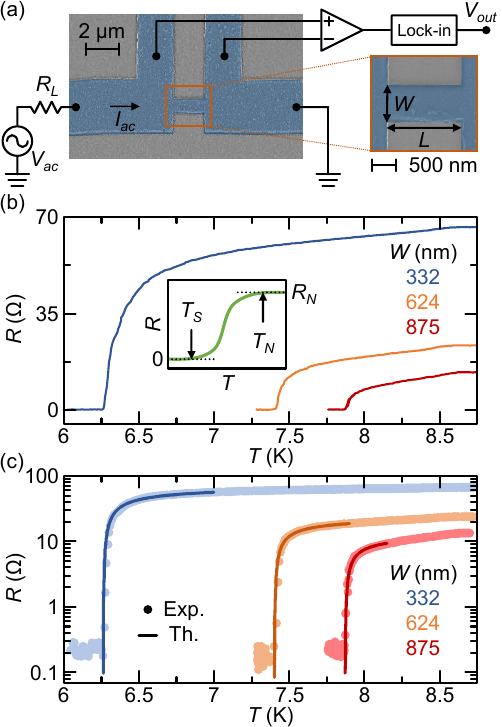}
\caption{\label{Fig1} (a) False-color scanning electron micrograph of a typical Nb nanowire. The devices are AC current-biased (amplitude $I_{ac}$) and the voltage drop across the wire ($V_{out}$) is measured via a voltage pre-amplifier connected
to a lock-in amplifier. $R_L=1$ M$\Omega$ serves as load resistor.
Inset: blow-up of the core of the device showing the Nb nanowire of length $L$ and width $W$.
(b) Resistance ($R$) versus temperature ($T$) characteristics of Nb nanowires of length $L=1.89\;\mu$m and thickness $t=49$ nm and widths ($W$) 332 nm (blue), 624 nm (orange) and 875 nm (red), respectively, sputtered with an initial chamber pressure $p=5\times 10^{-8}$ mbar.
Inset: schematic $R$ vs $T$ characteristic where the normal-state resistance ($R_N$), and the temperatures of fully superconducting ($T_S$) and fully normal ($T_N$) states are indicated.
(c) Logarithmic-scale representation of the experimental $R$ vs $T$ characteristics from panel (b) [dots, same of panel (b)] along with the $BKT$ model for dirty 2D superconductors (lines). The applied parameters are $R_{0}=80,\; 27,\; 15.1\; \Omega$, $b=0.12,\; 0.095,\; 0.09$ and $T_{BKT}=6.26,\; 7.4,\;7.87$ K for $W=332,\;624,\;875$ nm, respectively.}
\end{figure}

\section{Results and Discussion}

The devices were fabricated on an intrinsic silicon substrate covered with 300-nm-thick dry thermal SiO$_2$ layer by electron-beam lithography (EBL) of a poly(methyl methacrylate) (PMMA) mask, sputtering deposition of Niobium and final lift-off in acetone. The DC sputtering of Nb was performed at 150 W in a vacuum chamber with an initial pressure ($p$) $5\times 10^{-8}$ mbar or $8.8\times 10^{-8}$ mbar by flowing 10 sccm of Ar to maintain an operation pressure $p_{op}=5\times 10^{-4}$ mbar. The sputtering time ranged from 47 s to 109 s depending on the thickness of the Nb film.  The deposition was preceded by a 50 W pre-sputtering process at 50 sccm of Ar for 5 minutes.
Figure \ref{Fig1}(a) shows the false color scanning electron microscope (SEM) image of a typical Nb nanowire of length $L$ and width $W$. The measurements were performed with standard zero-bias 4-terminal lock-in technique by applying a voltage $V_{ac}$ of frequency 17 Hz on a load resistor $R_L=1$ M$\Omega\gg R$ (with $R$ the resistance of the device) to generate $I_{ac}=100$ nA. The voltage drop across the device ($V_{out}$) was recorded by a truly differential low-noise voltage pre-amplifier connected to a lock-in amplifier. The value of $I_{ac}$ was chosen to be much lower than the nanowires critical current, thus avoiding any lowering of the superconducting transition temperature due to sizeable current bias promoting dephasing.

Figure \ref{Fig1}(b) shows the dependence of $R$ on temperature ($T$) for three devices of thickness $t=49$ nm and different values of $W$ sputtered with an initial pressure $p=5\times 10^{-8}$ mbar. As expected, the normal-state resistance of the samples ($R_N$) increases by decreasing $W$, since all the devices share the length ($L=1.89\;\mu$m). Interestingly, the transition temperature to fully superconducting state ($T_S$) moves towards lower values in narrower devices. This shift is often related to the confinement of the wavefunction into a wire of lateral dimensions comparable to the superconducting coherence length ($\xi$). Indeed, one dimensional superconducting wires are usually described through the tilted washboard potential of Josephson junctions \cite{Barone1982,Ivanchenko1969} with the transition to the normal state due to thermally activated \cite{Altomare2006,Bezryadin2000,Giordano1988,Paolucci2020} or quantum \cite{Lehtinen2012} phase slips.

In our diffusive amorphous Nb devices, the superconducting coherence length takes the form $\xi=\sqrt{\hbar\,L/e^2N_FR_NW\,t\,\Delta_{0}}$, where $\hbar$ is the reduced Planck constant, $e$ is the electron charge, $N_F=7.34\times10^{47}$ J$^{-1}$m$^{-3}$ is the density of states at the Fermi level of Nb and $\Delta_{0}=1.764k_BT_S$ [with $k_B$ the Boltzmann constant and $T_S$ the temperature of full superconducting transition defined as $R(T_S)=10^{-3}R_N$, see the inset of Fig. \ref{Fig1}(b)] is the zero-temperature superconducting energy gap. The use of $T_S$ in the calculation of $\Delta_0$ provides the minimum value of the superconducting energy gap of the Nb film, which corresponds to the narrowest portion of the nanowire. To infer whether our devices are one dimensional, we calculate the maximum value of the coherence length $\xi_{max}\simeq8$ nm by substituting the experimental data obtained for the device with minimum $W=332$ nm ($L=1.89\;\mu$m, $R_N=66.3\;\Omega$) and lowest transition temperature $T_S=6.26$ K. Since $W\gg\xi_{max}$ and $t\gtrsim\xi_{max}$, all our Nb devices do not lie in the one dimensional wire case, but they are compatible with the two dimensional superconductor limit. The temperature dependence of the resistance of a dirty two-dimensional superconductor near its critical temperature can be calculated by the Berezinskii-Kosterlitz-Thouless ($BKT$) model. The latter takes the form \cite{Kosterlitz1973,Kosterlitz1974,Halperin1979,Yu2022}
\begin{equation}
 R(T)=R_0\, e^{-b\sqrt{\frac{T_{BKT}}{T}}},
 \label{Eq_BKT}
\end{equation}
where $R_0$ is the effective $BKT$ resistance, $b$ is a parameter related to the vortex core energy and $T_{BKT}$ is the $BKT$ transition temperature. The $BKT$ model shows an excellent agreement with our experimental data, as shown by Fig. \ref{Fig1}(c). This proves that our devices behave as two-dimensional superconductors, thus confirming that wavefunction confinement cannot account for the damping of the superconducting properties of the nanowires for decreasing $W$.

\begin{figure}
\centering
\includegraphics[width=0.95\columnwidth]{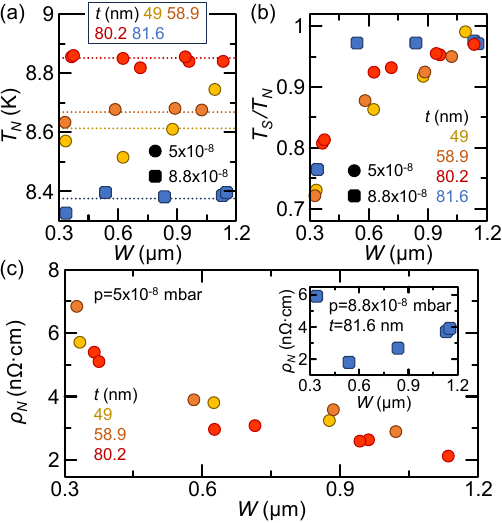}
\caption{\label{Fig2} (a) $T_N$ versus $W$ measured for samples of different thickness ($t$). Dots and squares indicate devices sputtered at a base pressure ($p$) $5\times 10^{-8}$ mbar and $8.8\times 10^{-8}$ mbar, respectively. The dotted lines show the average value of $T_N$ for each film thickness.
(b) $T_S/T_N$ versus $W$ measured for different values of $t$ and $p$.
(c) Normal-state resistivity ($\rho_N$) measured at 9 K versus $W$ for samples of different $t$ and $p=5\times 10^{-8}$ mbar. Inset: $\rho_N$ vs $W$ for a device of $t=81.6$ nm and $p=8.8\times 10^{-8}$ mbar.}
\end{figure}

To unveil the mechanism behind the weakening of the superconducting properties of our Nb devices with $W$, we will analyze in detail the $W$ dependence of the onset temperature for fully normal-state [$T_N$ defined as $R(T_N)=0.999R_N$, see the inset of Fig. \ref{Fig1}(b) for the definition], $T_S$ and the low-temperature normal-state resistivity ($\rho_N$). Figure \ref{Fig2}(a) shows the dependence of $T_N$ on $W$ for Nb films of different values of $t$ and $p$. First, $T_N$ is independent from $W$ for all the devices, thus indicating that there are portions of the samples showing the same superconducting correlations as the bare Nb film. Second, $T_N$ decreases by lowering $t$ at a given value of $p$, due to inverse proximity effect from the native partial oxidation of the film surface (of lower critical temperature) and localization effects \cite{Minhaj1994}. Third, $T_N$ decreases by sputtering at a larger value of $p$ for a given thickness, since the residual oxygen in the chamber accumulates at the Nb grain boundaries forming Niobium monoxide (NbO) and decreasing the film critical temperature \cite{Manzo2024,Kock1974,DeSorbo1963}. Meanwhile, $T_S$ monotonically decreases with $W$ [see Fig. \ref{Fig2}(b)]. In particular, all the samples grown at $p=5\times 10^{-8}$ mbar show a universal dependence of the ratio $T_S/T_N$ on the width independently from the value of $t$ (circles), thus indicating a common origin. This behavior is consistent with oxygen penetration from the PMMA mask, which serves as a reservoir \cite{Semenov2021-wt} and allows oxygen to diffuse into the Nb film during sputtering. By decreasing $W$, the lateral surface/volume ratio of the devices increases, thus the same amount of diffused oxygen entails a larger effect on the transport properties of the Nb sample. This process is also confirmed by the $T_S$ vs $W$ characteristics of the sample grown at $p=8.8\times 10^{-8}$ mbar [squares in Fig. \ref{Fig2}(b)]. Indeed, the larger initial concentration of interstitial oxygen implies that a larger lateral surface/volume ratio is necessary to observe a sizable effect on $T_S$, thus only the narrower device ($W=332$ nm) shows a suppression of the superconducting transition temperature.

In full agreement with previous data, the normal-state resistivity ($\rho_N$) of the devices sputtered at $p=5\times 10^{-8}$ mbar monotonically increases by lowering $W$, as shown in Fig. \ref{Fig2}(c). In fact, the increase of the amount of interstitial oxygen rises the number of scattering centers in the Nb film thus raising the device resistivity \cite{Manzo2024,Kock1974}. Differently, the samples sputtered at higher initial pressure show a larger $\rho_N$ at $W\sim1\;\mu$m which increases only for ultra-narrow devices $W=336$ nm [see the inset of Fig. \ref{Fig2}(c)]. This phenomenology confirms that the higher bulk content of oxygen of the Nb film conceals the diffusion from the PMMA surface.

\begin{figure}
\centering
\includegraphics[width=0.95\columnwidth]{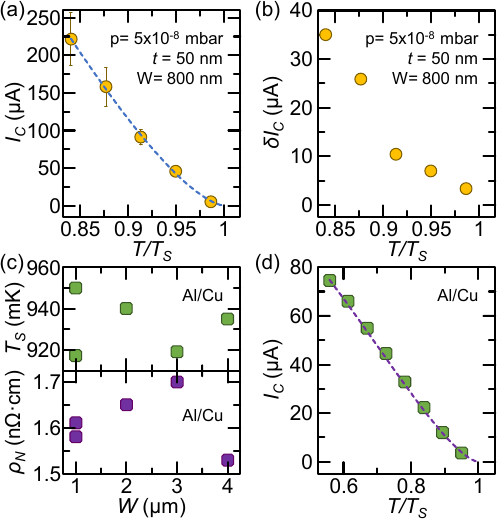}
\caption{\label{Fig5} (a) Critical current ($I_C$) versus normalized temperature ($T/T_S$) measured in a Nb device of $t=50$ nm, $W=800$ nm sputtered at a base pressure $p=5\times 10^{-8}$ mbar. The blue dashed line is the fit with the Bardeen curve providing $I_{C,0}\simeq1.39$ mA and $T_S=6.85$ K. (b) Spread of the measured values of critical current ($\delta I_C$) versus $T/T_S$ for the data in panel (a).
(c) $T_S$ (top, green squares) and $\rho_N$ (bottom, purple squares) versus $W$ measured for an Al/Cu bilayer with $t_{Al}=15$ nm and $t_{Cu}=15$ nm.
(d)  Critical current ($I_C$) versus normalized temperature ($T/T_S$) measured in a Al/Cu device of $t_{Al}=15$ nm, $t_{Cu}=15$ nm and $W=1\;\mu$m. The purple dashed line is the fit with the Bardeen curve providing $I_{C,0}\simeq131\;\mu$A and $T_S=895$ mK.}
\end{figure}

A possible source of tuning of the superconducting properties of nanodevices with the geometry is current crowding \cite{Hortensius2012}. In particular,  the condition to have maximum impact of current crowding is $\xi < W < \Lambda$ \cite{Hortensius2012}, where the Pearl length takes the form $\Lambda=2 \lambda_L^2/t$. Here, the London penetration depth for a diffusive superconductor can be written as $\lambda_L=\sqrt{\hbar\, R_N W\,t/\pi\,\mu_0\,L\,\Delta_{0}}$, with $\mu_0$ the magnetic permeability of vacuum. To investigate the role of current crowding we fabricated a device (with $t= 50$ nm and $W=800$ nm) showing $\xi_{800}\simeq6$ nm, $\lambda_{L,800}\simeq240$ and $\Lambda_{800}\simeq2.3\;\mu$m, thus respecting the above condition for strong current crowding. Figure \ref{Fig5}(a) shows the temperature dependence of the critical current ($I_C$) of this Nb device. The value of $I_C$ grows quickly by decreasing temperature. In particular, $I_C(0.9T_C)\simeq100\;\mu$A is 3 orders of magnitude larger than the bias current used in the $R$ vs. $T$ experiments ($|I_{ac}|=100$ nA), thus both movement towards lower temperatures and broadening of the superconducting transition are not compatible with current crowding. In addition, nanodevices affected by current crowding shown low dependence of $I_C$ near the superconducting transition \cite{Hortensius2012}. Instead, our experimental data show good agreement with the Bardeen equation for BCS superconductors $I_C(T)=I_{C,0}[1-(T/T_S)^2]^{3/2}$ (with $I_{C,0}$ the zero-temperature critical current) \cite{Bardeen1962}, as shown by the blue dashed line in Figure \ref{Fig5}(a). Furthermore, the variability of the critical current ($\delta I_C$) increases by rising (lowering) $I_C$ ($T$), as expected in BCS superconductors and in contrast with current crowding \cite{Hortensius2012}. 
Finally, we stress that the steadiness of $T_S/T_N$ for $W\geq600$ nm shown by the devices fabricated at $p=8.8\times 10^{-8}$ excludes further a possible role of current crowding for the behavior of samples sputtered at $p=5\times 10^{-8}$ mbar [see Fig. \ref{Fig2}(b)], since all the nanowires share the geometry depicted in Fig. \ref{Fig1}(a).

As further verification, we investigated devices with the same geometry made of a lifted-off Al/Cu ($t_{Al}=t_{Cu}=15$ nm). The samples are designed to have similar proportions of the width, coherence length and Pearl length of the Nb based nanowires, thus maximizing the possible impact of current crowding \cite{Hortensius2012}. Indeed, the coherence length $\xi_{Al/Cu}\simeq38.5$ nm, the London penetration length $\lambda_{Al/Cu}\simeq440$ nm and the Pearl length $\Lambda_{Al/Cu}\simeq12.9\;\mu$m ensure to respect the relation $\xi_{Al/Cu}<W<\Lambda_{Al/Cu}$ for all the devices. Differently from the Nb sample, the transport properties in the superconducting and normal state of Al/Cu devices are not affected by their width, as shown in Fig. \ref{Fig5}(c) by the dependence of  $T_S$ (top) and $\rho_N$ (bottom) on $W$. Furthermore, the temperature dependence of $I_C$ for the narrower device ($W=1\;\mu$m, where current crowding might provide the largest effect) follows the Bardeen prediction \cite{Bardeen1962}, as shown in Fig. \ref{Fig5}(d). Concluding, current crowding can be excluded as source of deterioration of the superconducting properties of our lifted-off Nb nanowires, since the Al/Cu samples of same geometry do not show any decay of superconductivity by channel narrowing.

\begin{figure}
\centering
\includegraphics[width=0.95\columnwidth]{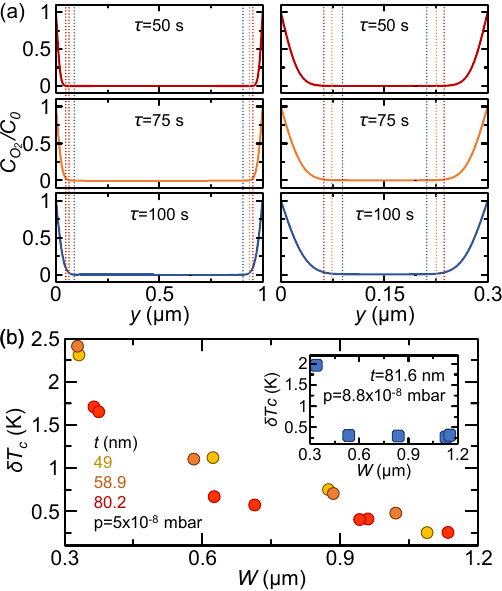}
\caption{\label{Fig3} (a) Normalized concentration of oxygen in the Nb film ($C_{ox}/C_{ox,max}$) versus the transverse position ($y$) calculated at $T=375$ K and different time periods ($\tau$) for device widths of 1 $\mu$m (left) and 300 nm (right). The vertical lines represent the positions corresponding to $C_{ox}=0$.
(b) Width of the superconducting transition ($\delta T_C=T_N-T_S$) versus $W$ measured for samples of different thickness ($t$) sputtered at a base pressure ($p$) $5\times 10^{-8}$ mbar. Inset: $\delta T_C$ vs $W$ for a device of $t=81.6$ nm and $p=8.8\times 10^{-8}$ mbar.}
\end{figure}

To account for our results, we exploit the one-dimensional Fick model describing the diffusion of oxygen in a Nb film \cite{bird2006transport}
\begin{equation}
\frac{C_{O_2}(y)}{C_{0}}=\text{erfc}\bigg( \frac{y}{2\sqrt{D_{O_2}\tau}}\bigg),
 \label{Eq_Fick}
\end{equation}
where $y$ is the direction perpendicular to the wire, $C_0$ is the oxygen content at the surface, $\tau$ is the growth time and $D_{O_2}=1.38\times10^{-2}\exp({-111530/R_GT})$ m$^{2}$s$^{-1}$ is the diffusion constant of oxygen in Nb [with $R_G\simeq8.31$ J/(K$\cdot$mol) the universal gas constant and $T$ the growth temperature] \cite{Hoerz1981}. Despite its simplicity, the  Fick model provides the order of magnitude of the penetration of oxygen from the PMMA surface in the Nb film during its growth. Figure \ref{Fig3}(a) shows the normalized transverse concentration profile of oxygen $C_{O_2}(y)/C_{0}$ for Nb devices 1 $\mu$m (left) and 300 nm (right) wide calculated at a sputtering temperature $T=375$ K and values of $\tau$ compatible with the experimental growth times. {At a given $\tau$, the oxygen diffusion has a stronger impact on narrow samples, since the increased oxygen concentration affects a larger relative portion of the nanowire.
This mechanism explains also the increased width of the superconducting transition, defined as $\delta T_C = T_N-T_S$, by decreasing $W$, as shown in Fig. \ref{Fig3}(b) for samples grown at $p=5\times 10^{-8}$ mbar. We note that the dependence of $\delta T_C$ on $W$ is largely unaffected by the device thickness. Differently, the samples sputtered at $p=8.8\times 10^{-8}$ mbar show an increase of $\delta T_C$ only for the smallest value of $W$ [see the inset of Fig. \ref{Fig3}(b)], since the diffusion from the PMMA walls does not increase sizably the oxygen concentration of devices of larger width. The strong temperature dependence of the Flick model (Eq. \ref{Eq_Fick}) excludes other sources of oxygen penetration and diffusion degrading the performance of our Nb devices. Indeed, to achieve the same lateral penetration at room temperature (T=293 K, the temperature of our fabrication facility) a time of $2.1\times 10^{6}$ s (about 24 days) would be necessary. Instead, the fabrication process of our devices is performed at room temperature within a few hours and the samples are kept under vacuum before the measurements.

The profile of $C_{O_2}(y)/C_{0}$ implies that $\rho_N$ is not constant along the nanowire cross-section. 
In full agreement with homogeneous Nb samples \cite{Manzo2024,Asada1969}, $T_S$ of our devices grown at $p=5\times 10^{-8}$ mbar shows the typical linear dependence on the effective normal-state resistivity $\rho_N$ (see  Fig. \ref{Fig4}).
It is evident that the slope of the $T_S(\rho_N)$ characteristics changes based on the film thickness. In particular, thinner devices show a steeper dependence of $T_S$ on $\rho_N$ [1.28 K/(n$\Omega\cdot$cm) for $t=49$ nm]. Indeed, the resistivity along the $y$ direction is more homogeneous for shorter deposition times, since the diffusion of oxygen into the Nb bulk from the PMMA is proportional to $\tau$ [see Fig. \ref{Fig3}(a)].
On the contrary, independently from the value of $t$, $T_S$ converges to $\sim8.5$ K for widths $W\geq1\;\mu$m. This indicates that wide samples acquire the same superconducting properties of a bare Nb film, thus confirming the role of oxygen diffusion from the PMMA lift-off mask. 

\begin{figure}
\centering
\includegraphics[width=0.95\columnwidth]{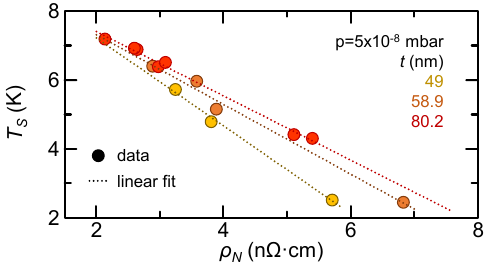}
\caption{\label{Fig4} $T_S$ versus $\rho_N$ for all the devices sputtered at a chamber base pressure $p=5\times 10^{-8}$ mbar. The dotted lines depict the linear fit of the data.}
\end{figure}

\section{Conclusions}

In summary, we have investigated the impact of sputtering base pressure, thickness ($t$) and width ($W$) on the superconducting properties of Nb nanowires fabricated by lift-off technique. Our results show, for the first time, the dependence of the superconducting properties of lifted-off Nb wires on their width. The resistance of our devices shows the typical $BKT$ temperature dependence of dirty two-dimensional superconductors \cite{Kosterlitz1973,Kosterlitz1974,Halperin1979,Yu2022}, thus excluding confinement of the superconducting order parameter and  phase slips \cite{Altomare2006,Bezryadin2000,Giordano1988,Paolucci2020,Lehtinen2012}.
At a given film thickness, our devices show a decrease of the fully superconducting-state transition temperature ($T_S$) by narrowing the wire, while the complete transition to the normal-state is almost unaffected. Consequently, the width of the superconducting transition ($\delta T_C$) rises by decreasing $W$. 
This phenomenology is attributed to oxygen diffusion from the electron-beam resist to the Nb thin film during the sputtering process, since other sources of oxidation would create homogeneous oxidation thus moving a steep superconducting transition towards lower temperatures. Further studies could exploit X-ray diffraction and atomic force microscopy, as already studied for non-lifted off Nb films with oxygen inclusions \cite{Ben_Shalom2016-ci,Thompson2017,Li_2018}. When fabricating wider superconducting electrodes is not feasible, a possible strategy to avoid the oxygen diffusion during the sputtering process is to deposit a thin film of another metal (or superconductor of lower critical temperature) before Nb which creates a barrier for oxygen. To avoid inverse proximity effect arising from the ohmic contact of Nb with a non-superconducting material \cite{MARTINIS2000}, the Nb film must be much thicker than the other metal providing the barrier for oxygen diffusion. 
Thus, our study provides the road map to exploit Nb in superconducting hybrid nanodevices \cite{Ben_Shalom2016-ci,Thompson2017,Li_2018} able to push quantum technology operating temperatures above 2 K, thus allowing the use of simple and cheap cryogenic systems.

\section*{Data Availability Statement}
The data that support the findings of this study are available from the corresponding author upon reasonable request.

\begin{acknowledgments}
The authors acknowledge A. Profeti for technical support.
F. B., F. P. and P. S. acknowledge partial financial support under the National Recovery and Resilience Plan (NRRP), Mission 4, Component 2, Investment 1.1, Call PRIN 2022 by the Italian Ministry of University and Research (MUR), funded by the European Union – NextGenerationEU – EQUATE Project, "Defect engineered graphene for electro-thermal quantum technology" - Grant Assignment Decree No. 2022Z7RHRS.
F. P. acknowledges the Italian Ministry of University and Research under the FIS2
Call - Grant No. FIS2023-00227 - QuLEAP, and the CSN V of INFN under the technology innovation grant STEEP for partial financial support.
C. P. acknowledges the PNRR MUR project PE0000023-NQSTI for partial financial support. 
\end{acknowledgments}

\providecommand{\noopsort}[1]{}\providecommand{\singleletter}[1]{#1}%

\end{document}